\documentclass[12pt,preprint]{aastex}

\usepackage{graphicx}
\usepackage{txfonts}
\newcommand{\rhop}{\rho_{\rm p}}
\newcommand{\rhoc}{\rho_{\rm c}}
\newcommand{\mutildep}{\tilde{\mu}_{\rm p}}
 \newcommand{\etac}{\eta_{\rm C}}
 \newcommand{\gammaAp}{\Gamma_{\rm Ap}}
\newcommand{\vap}{v_{\mathrm{Ap}}}
\newcommand{\vac}{v_{\mathrm{Ac}}}
\newcommand{\ld}{L_{\rm D}}
\newcommand{\ck}{c_{\rm k}}
\newcommand{\etacb}{\bar{\eta}_{\rm C}}
\newcommand{\ldlam}{L_{\rm D}/\lambda}
\newcommand{\tai}{\tau_{\rm Ap}}

\begin{document}

\title{SPATIAL DAMPING OF PROPAGATING KINK WAVES IN PROMINENCE THREADS}

	\shorttitle{Kink waves in prominence threads}

   \author{R. Soler,  R. Oliver, and J. L. Ballester}

   \affil{Departament de F\'isica, Universitat de les Illes Balears,
              E-07122, Palma de Mallorca, Spain\\
              \email{roberto.soler@uib.es}
             }

  \begin{abstract}
Transverse oscillations and propagating waves are frequently observed in threads of solar prominences/filaments and have been interpreted as kink magnetohydrodynamic (MHD) modes. We investigate the spatial damping of propagating kink MHD waves in transversely nonuniform and partially ionized prominence threads. Resonant absorption and ion-neutral collisions (Cowling's diffusion) are the damping mechanisms taken into account. The dispersion relation of resonant kink waves in a partially ionized magnetic flux tube is numerically solved by considering prominence conditions. Analytical expressions of the wavelength and damping length as functions of the kink mode frequency are obtained in the Thin Tube and Thin Boundary approximations. For typically reported periods of thread oscillations, resonant absorption is an efficient mechanism for the kink mode spatial damping, while ion-neutral collisions have a minor role. Cowling's diffusion dominates both the propagation and damping for periods much shorter than those observed. Resonant absorption may explain the observed spatial damping of kink waves in prominence threads. The transverse inhomogeneity length scale of the threads can be estimated by comparing the observed wavelengths and damping lengths with the theoretically predicted values. However, the ignorance of the form of the density profile in the transversely nonuniform layer introduces inaccuracies in the determination of the inhomogeneity length scale.
\end{abstract}


   \maketitle
%

%________________________________________________________________

\section{INTRODUCTION}

Waves and oscillatory motions are frequently reported in the observations of solar prominences and filaments \citep[see reviews by][]{oliverballester02,ballester,engvold08,mackay}. In high-resolution observations, the prominence fine structures (threads) often display transverse oscillations of small amplitude \citep[e.g.,][]{lin05,lin07,lin09,okamoto,ning}, which have been interpreted as kink magnetohydrodynamic (MHD) waves \citep[e.g.,][]{diaz2002,dymovaruderman,hinode,lin09}. The observed threads in H$\alpha$ images are between 3,000~km and 28,000~km long, and between 100~km and 600~km wide \citep{lin04,lin08}. The threads outline part of much larger magnetic flux tubes which are probably rooted in the solar photosphere. The majority of observed periods of transverse thread oscillations roughly range between 1~min and 10~min, but a few detections of longer periods of about 20~min have been also informed \citep[e.g.,][]{yi,lin04}. The wavelengths are usually between 700~km and 8,000~km, although values up to 250,000~km have been reported \citep{okamoto}. Recently, \citet{solerfine} pointed out that the short periods and wavelengths are consistent with an interpretation in terms of propagating waves, while periods larger than 10~min and wavelengths longer than 100,000~km could correspond to standing oscillations of the whole magnetic tube. In the case of standing oscillations, the value of the wavelength is not strongly influenced by the thread properties but is mainly determined by the total length of the magnetic tube, since the fundamental kink mode wavelength is twice the length of the tube, approximately \citep[see details in][]{solerfine}. Although there are no direct measurements of the length of prominence magnetic tubes, this parameter is estimated around 10$^5$~km. This rough estimation is in agreement with the wavelengths reported by \citet{okamoto}. In addition, a common feature of the observations is that the oscillations are strongly damped \citep[e.g.,][]{terradasobs,ning,lin09}.

Motivated by the observational evidence, great effort has been recently devoted to the theoretical study of both temporal and spatial damping of MHD waves in prominence plasmas. Temporal damping is investigated for waves with fixed wavelength, while spatial damping is studied for propagating waves with fixed frequency. Both phenomena have been extensively investigated in unbounded and homogeneous prominence plasmas by assuming different damping mechanisms \citep[e.g.,][]{carbonell04,carbonell06,carbonell10,forteza07,forteza08}. The reader is referred to \citet{oliver}, \citet{arreguiballester}, and references therein for a complete account of the theoretical works. 

In the case of prominence thread oscillations, works so far have focused on temporal damping by mechanisms as, e.g., non-adiabatic effects \citep{solernonad}, ion-neutral collisions \citep{solerPI}, and resonant absorption \citep{arregui08,arregui10,solerslow,solerRAPI,solerfine}. The conclusions of these works indicate that resonant absorption is efficient enough to provide realistic kink mode damping times consistent with the reported strong damping, whereas non-adiabatic effects are negligible and ion-neutral collisions are only important for shorter wavelengths than those observed. In the case of spatial damping, \citet{pecseli} studied the effect of ion-neutral collisions but restricted themselves to Alfv\'en waves and kink modes were not investigated. Although spatial damping of kink waves has been studied in the context of coronal loops \citep[e.g.,][]{pascoe,spatialterradas}, to our knowledge no detailed investigation taking into account the peculiar properties of prominence threads can be found in the existing literature. The purpose of this paper is to fill this gap in the literature, as the recent observations of wave damping in solar prominences need to be understood.

Here, we study the spatial damping of propagating kink waves in prominence threads. Our model is composed of a cylindrical magnetic flux tube with partially ionized prominence plasma, representing a thread, surrounded by a fully ionized coronal environment. The thread is non-uniform in the transverse direction. Resonant absorption and ion-neutral collisions are assumed as the damping mechanisms. We use the $\beta = 0$ approximation, with $\beta$ the ratio of the gas pressure to the magnetic pressure, and the Thin Boundary approach to describe the effect of resonant absorption in the Alfv\'en continuum using the connection formulas for the perturbations across the resonant layer \citep[e.g.,][]{sakurai,goossens92}. We determine the dominant damping mechanism and obtain analytical expressions for the wavelength, the damping length, and their ratio as functions of the kink mode frequency.

This paper is organized as follows. Section~\ref{sec:equations} contains the description of the model configuration and the basic Equations. First, the problem is attacked analytically in Section~\ref{sec:analytical} by adopting the thin tube approximation. Later on, the full dispersion relation is numerically solved and a parametric study of the wavelength and damping length of the kink mode as functions of the period is performed in Section~\ref{sec:numerical}. Finally, the conclusions of this work are given in Section~\ref{sec:conclusions}.

%__________________________________________________________________

 \section{MODEL AND DISPERSION RELATION} 
 \label{sec:equations}

 \begin{figure}[!htp]
\centering
\epsscale{0.75}
\plotone{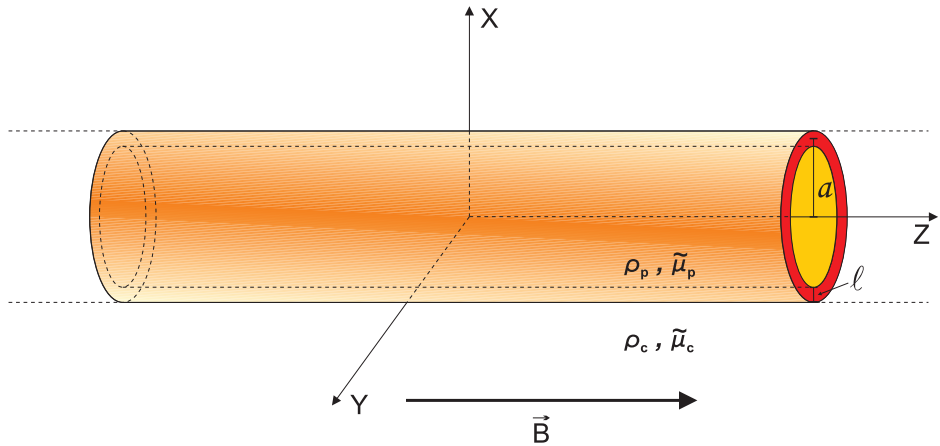}
\caption{Sketch of the prominence thread model adopted in this work. \label{fig:equilibrium}}
\end{figure}

The equilibrium configuration is composed of a straight magnetic cylinder of radius $a$ embedded in a homogeneous environment representing the coronal medium (see Figure~\ref{fig:equilibrium}). We use cylindrical coordinates, namely $r$, $\varphi$, and $z$ for the radial, azimuthal, and longitudinal coordinates, respectively. The magnetic field is uniform and along the axis of the cylinder, ${\bf B} = B \hat{e}_z$, with $B$ constant everywhere. Hereafter, subscripts p and c denote prominence and coronal quantities, respectively. The density within the prominence thread is denoted by $\rhop$, while the coronal density is $\rhoc$. Both $\rhop$ and $\rhoc$ are homogeneous. A transverse transitional layer is included in the radial direction, where the density varies continuously between the internal and external densities. We do not specify the form of the density profile at this stage. The transverse inhomogeneous length scale in the transitional layer is given by the radio $l/a$, with $l$ the thickness of the layer. This ratio ranges from $l/a = 0$ if no transitional layer is present, to $l/a = 2$ if the whole tube is radially inhomogeneous. Due to the presence of the transverse transitional layer, the kink mode is resonantly coupled to Alfv\'en continuum modes. The resonance leads to the kink mode damping as the energy is transferred to Alfv\'en modes at the Alfv\'en resonance position. This mechanism is known as resonant absorption.

The prominence plasma is partially ionized and we adopt the single-fluid formalism \citep[e.g.,][]{brag}. The ionization degree is arbitrary and is denoted here by the mean atomic weight of the prominence material, $\mutildep$. This parameter takes values in the range $0.5 \leq \mutildep \leq 1$, where $\mutildep=0.5$ corresponds to a fully ionized plasma and $\mutildep=1$ to a fully neutral gas \citep[see details in, e.g.,][]{forteza07,solerPI}. The external coronal medium is assumed fully ionized. The basic MHD Equations governing a partially ionized plasma can be found in, e.g., \citet{forteza07,pinto,solerphd}. By assuming the $\beta = 0$ approximation and linear perturbations from the equilibrium state, the basic MHD Equations discussed in this work are
 \begin{equation}
 \rho \frac{\partial {\bf v}}{\partial t} = \frac{1}{\mu}\left( \nabla \times {\bf b} \right) \times {\bf B}, \label{eq:motionbeta0app} 
\end{equation}
\begin{equation}
  \frac{\partial {\bf b}}{\partial t} =  \nabla \times \left( {\bf v} \times {\bf B}\right) + \nabla \times \left\{ \frac{\etac}{B^2} \left[ \left( \nabla \times {\bf b} \right) \times {\bf B} \right] \times {\bf B} \right\}, \label{eq:inductionapp}
\end{equation}
where $\rho$ is the local density, and ${\bf v} = \left( v_r, v_\varphi, v_z \right)$ and ${\bf b} = \left( b_r, b_\varphi, b_z \right)$ are the velocity and the magnetic field perturbations, respectively. Note that $v_z = 0$ in the $\beta = 0$ approximation. In a partially ionized plasma, the induction equation contains a term accounting for Cowling's diffusion, i.e., the second term on the right-hand side of Equation~(\ref{eq:inductionapp}). Cowling's diffusion represents enhanced magnetic diffusion caused by ion-neutral collisions, which is several orders of magnitude more efficient than classical Ohm's diffusion and is the dominant effect in partially ionized plasmas \citep{cowling}. For this reason, here we neglect Ohm's diffusion and other terms of minor importance present in the generalized induction equation \citep[see, e.g., Equation~(14) of][]{forteza07}. Cowling's diffusion coefficient, $\etac$, depends on the ionization degree through $\mutildep$ as well as on the plasma physical conditions. The expression of $\etac$ for a hydrogen plasma can be found in, e.g., \citet{pinto} and  \citet{solerPI}, whereas for a plasma composed of hydrogen and helium see \citet{solerhelium}. As the effect of helium is negligible for realistic helium abundances in prominences, here we consider a pure hydrogen plasma. The effect of Cowling's diffusion is neglected in the external medium because the corona is assumed fully ionized.

We follow an approach based on normal modes. Since $\varphi$ and $z$ are an ignorable coordinates, the perturbations are expressed proportional to $\exp \left(i m \varphi + i k_z z - i \omega t  \right)$, where $\omega$ is the oscillatory frequency, $k_z$ is the longitudinal wavenumber, and $m$ is the azimuthal wavenumber ($m=1$ for the kink mode). Alternatively, the problem could be investigated by means of time-dependent simulations of driven waves as in \citet{pascoe}. However, in the linear regime the different values of $m$ and $k_z$ are decoupled from each other, and a normal mode analysis is a simpler procedure for linear waves. If Cowling's diffusion is neglected, our configuration corresponds to that studied by \citet{spatialterradas} for propagating kink waves in coronal loops. We extend their investigation by incorporating the effect of Cowling's diffusion due to ion-neutral collisions

By using the Thin Boundary (TB) approach \citep[see details in, e.g.,][]{goossensrev,goossensiau}, the analytical dispersion relation for resonantly damped kink waves propagating in a transversely nonuniform and partially ionized prominence thread was obtained by \citet[Equation~(25)]{solerRAPI}. If partial ionization is not considered and the effect of Cowling's diffusion is absent, the dispersion relation of \citet{solerRAPI} reduces to that investigated by \citet[Equation~(28)]{spatialterradas} for kink waves in coronal loops. \citet{solerRAPI} checked that the solutions of their dispersion relation are in excellent agreement with the solutions obtained from the full numerical integration of the MHD equations beyond the TB approximation. Therefore, the dispersion relation derived by \citet{solerRAPI} correctly describes the kink mode behavior in our model and complicated numerical integrations are not needed. The dispersion relation obtained by \citet{solerRAPI} in the case of a straight and homogeneous magnetic field is
\begin{equation}
\frac{n_c}{\rhoc \left( \omega^2 -  k_z^2 \vac^2 \right)} \frac{K'_m \left( n_ {\rm c} a \right)}{K_m \left( n_{\rm c} a \right)} - \frac{m_{\rm p}}{\rhop \left( \omega^2 - k_z^2 \gammaAp^2 \right)} \frac{J'_m \left( m_{\rm p} a \right)}{J_m \left( m_{\rm p} a \right)} = - i \pi \frac{m^2 / r_{\rm A}^2}{\omega^2 \left| \partial_r \rho  \right|_{r_{\rm A}}}, \label{eq:reldisper}
\end{equation}
where $J_m$ and $K_m$ are the Bessel function and the modified Bessel function of the first kind of order $m$ \citep{abra}, respectively, and the quantities $m_{\rm p}$ and $n_{\rm c}$ are defined as
\begin{equation}
 m_{\rm p}^2 = \frac{\left( \omega^2 - k_z^2 \gammaAp^2  \right)}{\gammaAp^2}, \qquad n_{\rm c}^2 = \frac{\left( k_z^2 \vac^2 - \omega^2  \right)}{\vac^2},\label{eq:mn}
\end{equation}
where $\gammaAp^2 = \vap^2 - i \omega \etac$ is the modified prominence Alfv\'en speed squared \citep{forteza08}, with $\vap = \frac{B}{\sqrt{\mu \rhop}}$ and $\vac = \frac{B}{\sqrt{\mu \rhoc}}$ the prominence and coronal Alfv\'en speeds, respectively, and $\mu = 4 \pi \times 10^{-7}$~N~A$^{-2}$ the magnetic permeability. In addition, $r_{\rm A}$ is the Alfv\'en resonance position, and $\left| \partial_r \rho  \right|_{r_{\rm A}}$ is the radial derivative of the transverse density profile at the Alfv\'en resonance position.
 
\citet{solerRAPI} studied the temporal damping of kink waves, hence they assumed a fixed, real $k_z$ and solved Equation~(\ref{eq:reldisper}) to obtain the complex frequency. Here, we investigate the spatial damping and proceed the other way round, i.e., we fix a real $\omega$ and solve Equation~(\ref{eq:reldisper}) to obtain the complex wavenumber. Then, the period, $P$, wavelength, $\lambda$, and damping length, $\ld$, are computed as follows,
\begin{equation}
 P = \frac{2 \pi}{\omega}, \qquad \lambda = \frac{2 \pi}{k_{z\rm R}}, \qquad \ld = \frac{1}{k_{z\rm I}},
\end{equation}
with $k_{z\rm R}$ and $k_{z\rm I}$ the real and imaginary parts of $k_z$, respectively.

% %__________________________________________________________________
% 
\section{ANALYTICAL APPROXIMATIONS}
 \label{sec:analytical}

Some analytical progress can be performed before solving Equation~(\ref{eq:reldisper}) by means of numerical methods. To do so, we adopt the Thin Tube (TT) limit, i.e., $\lambda / a \gg 1$. A fist-order expansion of Equation~(\ref{eq:reldisper}) gives the dispersion relation in both the TT and TB approximations, namely
\begin{equation}
 \rhop \left( \omega^2 - k_z^2 \gammaAp^2 \right) + \rhoc \left( \omega^2 - k_z^2 \vac^2 \right) - i \pi \frac{m}{r_{\rm A}}\frac{\rhop \rhoc}{\left| \partial_r \rho  \right|_{r_{\rm A}}} \frac{\left( \omega^2 - k_z^2 \gammaAp^2 \right) \left( \omega^2 - k_z^2 \vac^2 \right)}{\omega^2} = 0. \label{eq:reltt}
\end{equation}
If both Cowling's diffusion and resonant absorption are omitted, the solution to Equation~(\ref{eq:reltt}) is
\begin{equation}
 k_z^2 = \frac{\omega^2}{\ck^2}, \label{eq:ideal}
\end{equation}
with $\ck^2 = \frac{2 B^2}{\mu \left( \rhop + \rhoc  \right)}$ the kink speed squared. Equation~(\ref{eq:ideal}) corresponds to the ideal, undamped kink mode. The solutions to Equation~(\ref{eq:reltt}) considering the different damping mechanisms are discussed next.

\subsection{Damping by Cowling's diffusion}

In the absence of transverse transitional layer, i.e., $l/a=0$, resonant absorption does not take place and the damping is  due to Cowling's diffusion exclusively. In such a case, the third term on the left-hand side of Equation~(\ref{eq:reltt}) is absent. We write the wavenumber as $k_z = k_{z\rm R} + i k_{z\rm I}$ and use Equation~(\ref{eq:reltt}) to obtain the exact expressions for $k_{z\rm R}^2$ and $k_{z\rm I}^2$, namely
\begin{eqnarray}
 k_{z\rm R}^2 &=& \frac{1}{2} \frac{\omega^2 \ck^2}{\ck^4 + \omega^2 \etacb^2} \left( \sqrt{1+ \frac{\omega^2 \etacb^2}{\ck^4}} + 1  \right), \label{eq:kzrcow1} \\
k_{z\rm I}^2 &=& \frac{1}{2} \frac{\omega^2 \ck^2}{\ck^4 + \omega^2 \etacb^2} \left( \sqrt{1+ \frac{\omega^2 \etacb^2}{\ck^4}} - 1  \right),\label{eq:kzicow1}
\end{eqnarray}
with $\etacb = \frac{\rhop}{\rhop+\rhoc}\etac$. By combining Equations~(\ref{eq:kzrcow1}) and (\ref{eq:kzicow1}), we compute the ratio of the damping length to the wavelength as
\begin{equation}
 \frac{\ld}{\lambda} = \frac{k_{z\rm R}}{2\pi k_{z\rm I}} = \frac{1}{2\pi} \frac{\ck^2 + \sqrt{\ck^4 + \omega^2 \etacb^2}}{\omega \etacb}. \label{eq:cowldlam1}
\end{equation}

Equations~(\ref{eq:kzrcow1})--(\ref{eq:cowldlam1}) are exact expressions that can be further simplified depending on the value of the ratio $\frac{\omega^2 \etacb^2}{\ck^4}$. For $\frac{\omega^2 \etacb^2}{\ck^4} \ll 1$, i.e., in the limit of low frequency ($\omega$ small) and/or large ionization degree ($\etacb$ small), Equations~(\ref{eq:kzrcow1})--(\ref{eq:cowldlam1}) simplify to
\begin{eqnarray}
 k_{z\rm R}^2 &\approx& \frac{\omega^2}{\ck^2 \left( 1 + \frac{\omega^2 \etacb^2}{\ck^4} \right)} \approx \frac{\omega^2}{\ck^2},\label{eq:kzrcow2} \\
k_{z\rm I}^2 &\approx& \frac{1}{4}\frac{\omega^4 \etacb^2}{\ck^6 \left( 1 + \frac{\omega^2 \etacb^2}{\ck^4} \right)} \approx \frac{1}{4} \frac{\omega^4 \etacb^2}{\ck^6}, \label{eq:kzicow2} \\
\frac{\ld}{\lambda} &\approx& \frac{1}{\pi} \frac{\ck^2}{\omega \etacb} \label{eq:cowldlam2}.
\end{eqnarray}
On the contrary, if $\frac{\omega^2 \etacb^2}{\ck^4} \gg 1$, i.e., high frequency and/or small ionization degree, the equivalent expressions are
\begin{eqnarray}
 k_{z\rm R}^2 &\approx& k_{z\rm I}^2 \approx \frac{1}{2} \frac{\omega}{\ck^2 \etacb}, \label{eq:cowdomi} \\
\frac{\ld}{\lambda} &\approx& \frac{1}{2\pi}. \label{eq:ldlamlimcow}
\end{eqnarray}
Thus, for $\frac{\omega^2 \etacb^2}{\ck^4} \ll 1$ the ratio of the damping length to the wavelength is inversely proportional to both $\omega$ and $\etacb$, and $k_{z\rm R}^2$ coincides with the ideal value (Equation~(\ref{eq:ideal})). This case corresponds to a weakly damped kink mode. On the other hand, for $\frac{\omega^2 \etacb^2}{\ck^4} \gg 1$, $\ldlam$ is independent of $\omega$ and $\etacb$  and the wave behavior is governed by diffusion. By assuming typical values for the parameters in the context of oscillating prominence threads, e.g., $P=3$~min, $B=5$~G, and $\rhop/\rhoc = 200$, we obtain $\frac{\omega^2 \etacb^2}{\ck^4} \approx 6\times 10^{-17}$ for $\mutildep = 0.5$ and $\frac{\omega^2 \etacb^2}{\ck^4} \approx 1.6 \times 10^{-4}$ for $\mutildep = 0.99$, meaning that the case $\frac{\omega^2 \etacb^2}{\ck^4} \ll 1$ is more realistic in the context of oscillating threads even for an almost neutral plasma.

\subsection{Damping by resonant absorption and Cowling's diffusion}

Next, we take the case $l/a \neq 0$ into account and study the combined effect of resonant absorption and Cowling's diffusion. The third term on the left-hand side of Equation~(\ref{eq:reltt}) is now present. As before, we write $k_z = k_{z\rm R} + i k_{z\rm I}$ and put this expression in Equation~(\ref{eq:reltt}). Since it is very difficult to give exact expressions for $k_{z\rm R}$ and $k_{z\rm I}$ in the general case, we focus on $\ldlam$ and restrict ourselves to $\frac{\omega^2 \etacb^2}{\ck^4} \ll 1$. Following the procedure of \citet{spatialterradas}, we assume weak damping, i.e., $k_{z\rm I} \ll k_{z\rm R}$, and neglect terms with $k_{z\rm I}^2$. The following process is long but straightforward, and we refer the reader to \citet{spatialterradas} for details. Finally, we arrive at the expression for the ratio of the damping length to the wavelength as
\begin{equation}
 \frac{\ld}{\lambda} \approx  \left( \pi \frac{\omega \etacb}{\ck^2} + \frac{m}{\mathcal{F}} \frac{l}{a} \frac{\rhop-\rhoc}{\rhop+\rhoc} \right)^{-1}, \label{eq:ldlamboth}
\end{equation}
where the first term within the parentheses accounts for Cowling's diffusion and the second term for resonant absorption. The factor $\mathcal{F}$ in the second term takes different values depending on the density profile within the inhomogeneous layer. For example, $\mathcal{F} = 4/\pi^2$ for a linear profile \citep{goossens2002}, while $\mathcal{F} = 2/\pi$ for a sinusoidal profile with $r_{\rm A} \approx a$ \citep{rudermanroberts}.  If the term related to  resonant absorption is absent, Equation~(\ref{eq:ldlamboth}) reverts to Equation~(\ref{eq:cowldlam2}). On the other hand, if the term related to Cowling's diffusion is dropped, Equation~(\ref{eq:ldlamboth}) coincides with Equation~(13) of \citet{spatialterradas}. 

The relative importance of the two terms in Equation~(\ref{eq:ldlamboth}) can be assessed by performing their ratio as
\begin{equation}
 \epsilon \equiv \frac{\left( \ldlam \right)_{\rm RA}}{\left( \ldlam \right)_{\rm C}} \approx \pi \mathcal{F} \frac{a}{l} \frac{ \omega \etacb}{\ck^2 m}  \frac{\rhop+\rhoc}{\rhop-\rhoc} = \pi \mathcal{F} \frac{a}{l} \frac{ \omega \etac}{\ck^2 m }  \frac{\rhop}{\rhop-\rhoc}, \label{eq:delta}
\end{equation}
where $\left( \ldlam \right)_{\rm RA}$ and $\left( \ldlam \right)_{\rm C}$ stand for the damping ratio by resonant absorption and Cowling's diffusion, respectively. By considering as before $P=3$~min, $B=5$~G, and $\rhop/\rhoc = 200$, and adopting a linear profile with $l/a = 0.2$, we obtain $\epsilon \approx 8 \times 10^{-8}$ for $\mutildep = 0.5$ and $\epsilon \approx 0.12$ for $\mutildep = 0.99$, meaning that in the TT limit resonant absorption dominates the kink mode spatial damping for typical parameters of thread oscillations. This result is equivalent to that obtained by \citet{solerRAPI} in the case of temporal damping.

\section{NUMERICAL RESULTS} 
\label{sec:numerical}

Now, we solve the dispersion relation (Equation~(\ref{eq:reldisper})) by means of standard numerical procedures. In the following figures, both the wavelength, $\lambda$, and the damping length, $\ld$, are plotted in dimensionless form with respect to the thread mean radius, $a$. The period, $P$, is computed in units of the internal Alfv\'en travel time, $\tai = a / \vap$. Unless otherwise stated, the results have been computed with $\rhop = 5 \times 10^{-11}$~kg~m$^{-3}$, $\rhop / \rhoc = 200$, and $B=5$~G. With these parameters, $\vap \approx 63$~km~s$^{-1}$ and, for $a=100$~km, $\tai \approx 1.59$~s.

\begin{figure}[!htp]
\centering
\epsscale{0.49}
\plotone{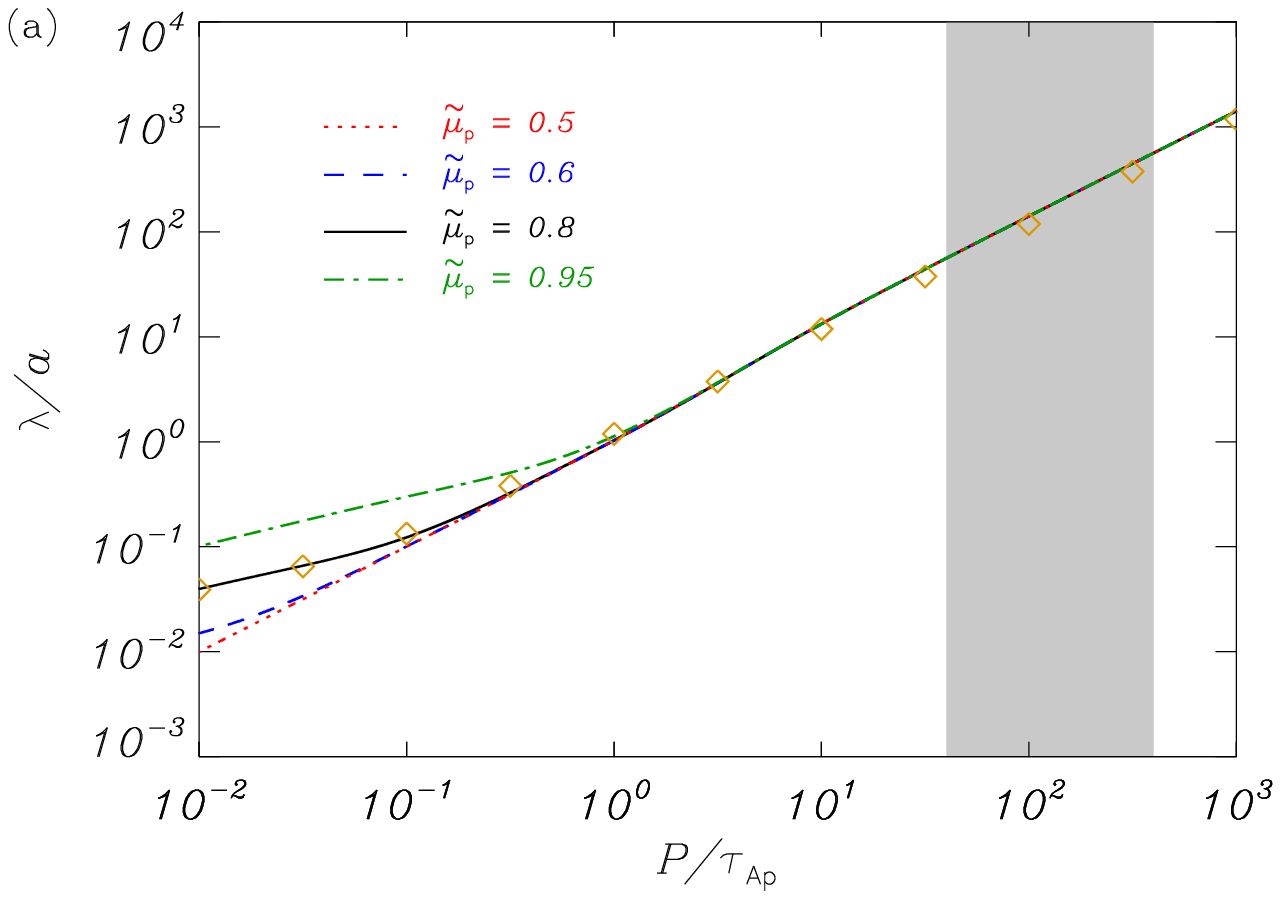}
\plotone{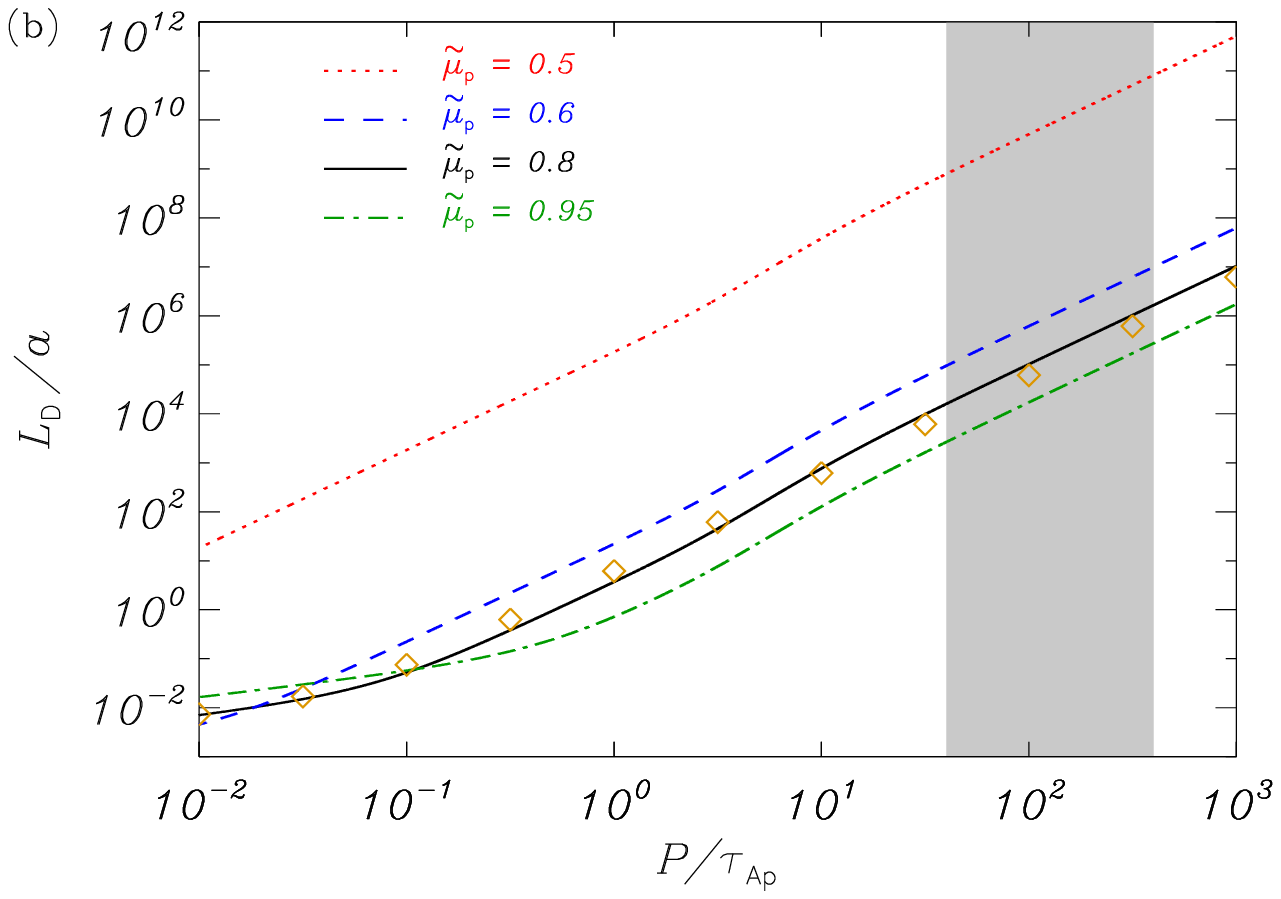}
\plotone{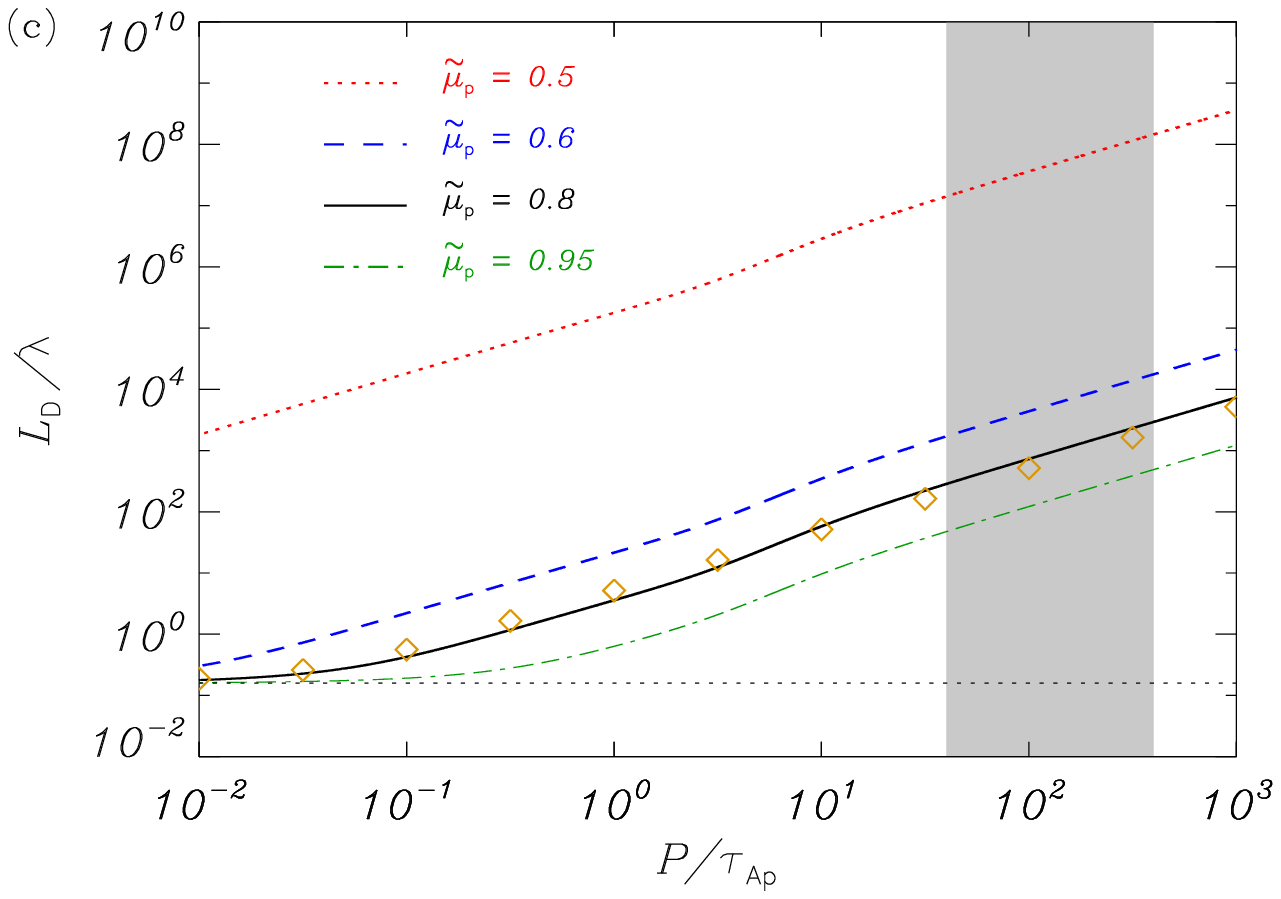}
\caption{Results for the kink mode spatial damping in the case $l/a=0$: (a) $\lambda/a$, (b) $\ld/a$, and (c) $\ldlam$ versus $P/\tai$ for $\mutildep=$~0.5, 0.6, 0.8, and 0.95. Symbols in panels (a), (b), and (c) correspond to the analytical solution in the TT approximation given by Equations~(\ref{eq:kzrcow1}), (\ref{eq:kzicow1}), and (\ref{eq:cowldlam1}), respectively, while the horizontal dotted line in panel (c) corresponds to the limit of $\ldlam$ for high frequencies (Equation~(\ref{eq:ldlamlimcow})). The shaded area denotes the range of observed periods of thread oscillations. \label{fig:cow}}
\end{figure}

Figure~\ref{fig:cow} displays $\lambda/a$, $\ld/a$, and $\ldlam$ versus $P/\tai$ for the case $l/a=0$, i.e., the damping is due to Cowling's diffusion exclusively. We compute the results for different values of $\mutildep$. The shaded areas in Figure~\ref{fig:cow} and in the other figures represent the range of observed periods of transverse thread oscillations, i.e., 1~min -- 10~min, corresponding to $40 \lesssim P/\tai \lesssim 400$, approximately. Regarding the wavelength, we see that the effect of Cowling's diffusion is only relevant for periods much shorted than those observed. This is in agreement with Equations~(\ref{eq:kzrcow2}) and (\ref{eq:cowdomi}). On the other hand, an almost neutral plasma, i.e., $\mutildep \to 1$, has to be considered to obtain an efficient damping and to achieve small values of $\ldlam$ within the relevant range of periods. Although we do not know the exact ionization degree in prominence threads, such very large values of $\mutildep$ are probably unrealistic \citep[see, e.g.,][]{labrosse}. The analytical expressions for $\lambda$, $\ld$, and $\ldlam$ in the TT case given by Equations~(\ref{eq:kzrcow1}), (\ref{eq:kzicow1}), and (\ref{eq:cowldlam1}), respectively, are in good agreement with the full results in the whole range of periods (see symbols in Figure~\ref{fig:cow}).

 \begin{figure}[!htp]
\centering
\epsscale{0.49}
\plotone{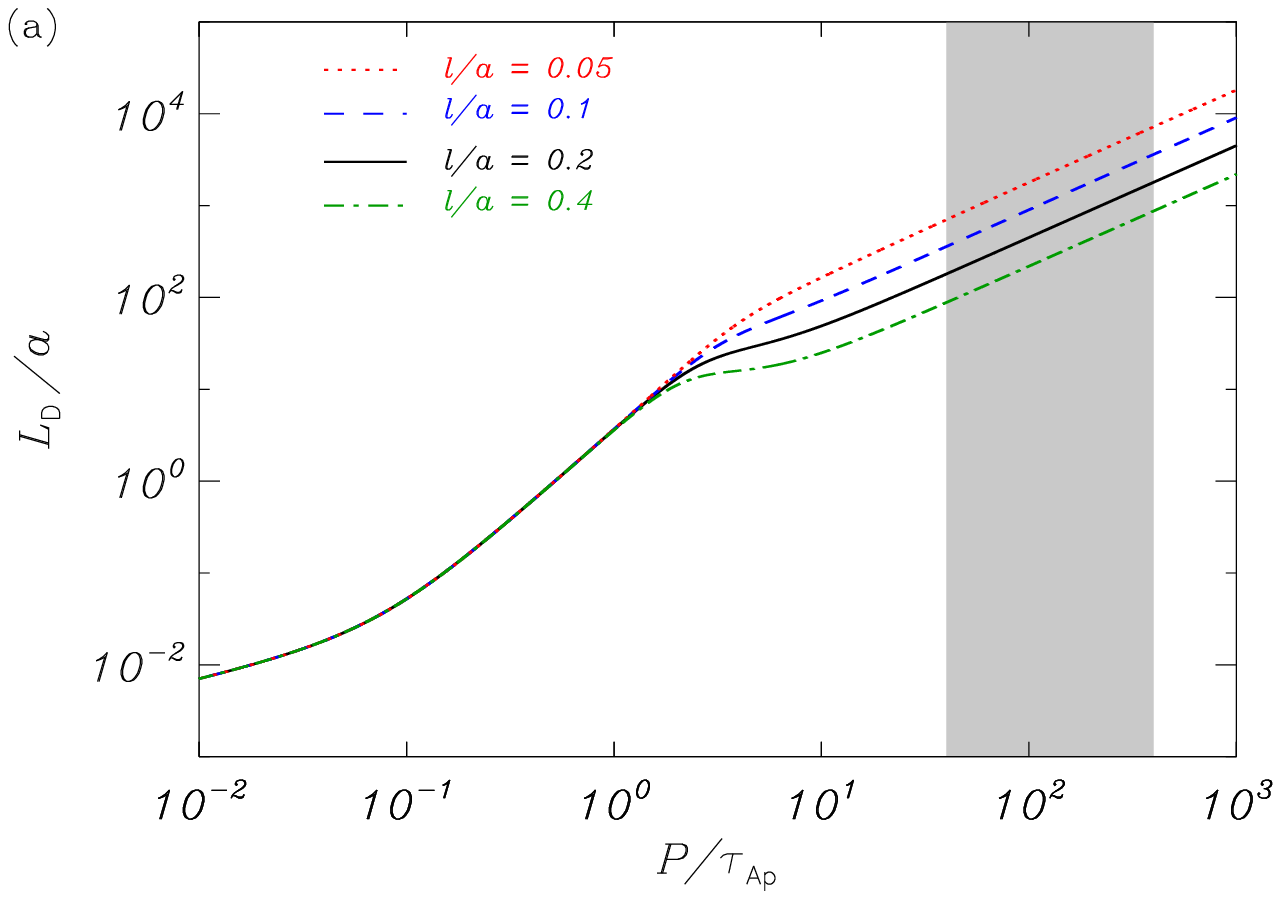}
\plotone{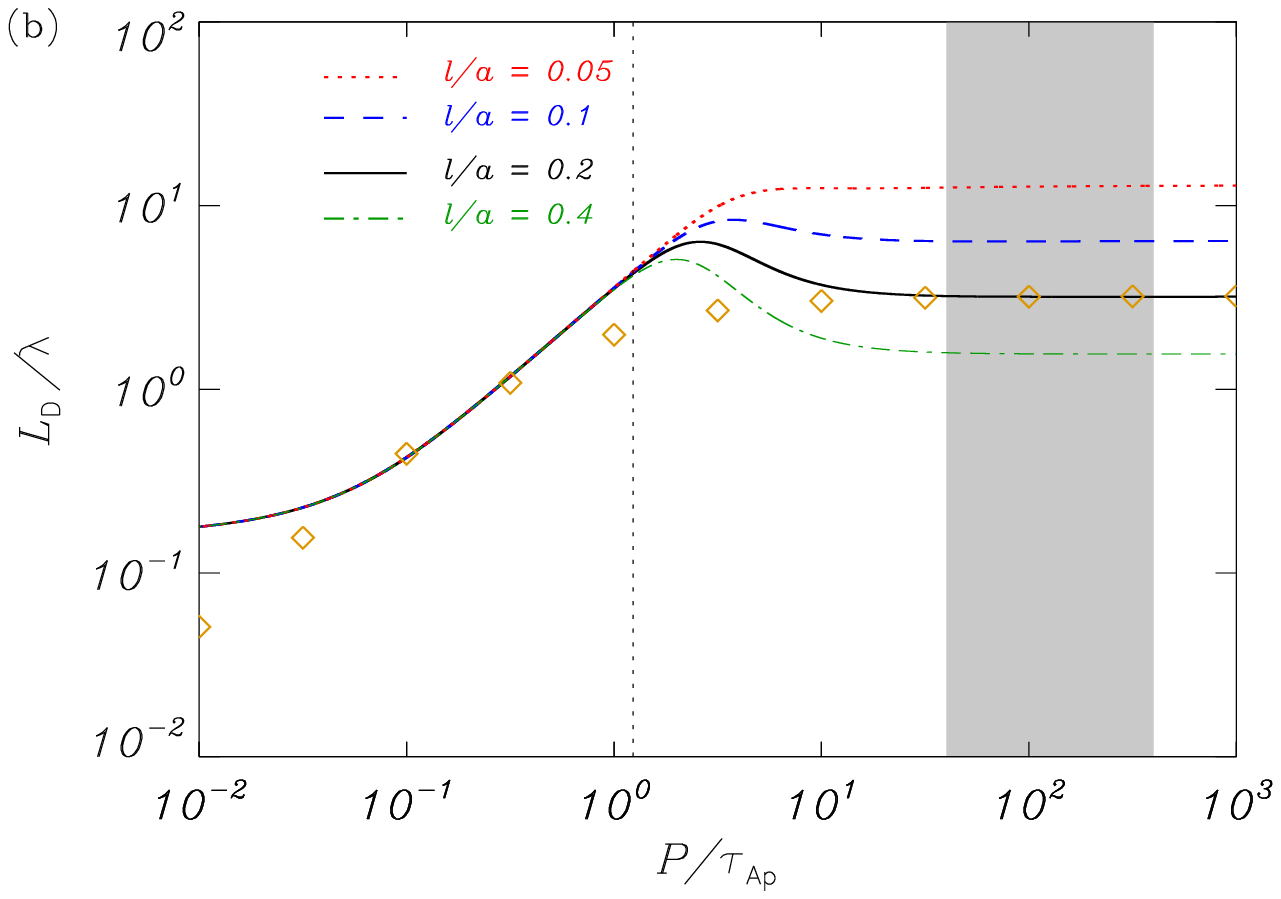}
\caption{Results for the kink mode spatial damping in the case $l/a \neq 0$: (a) $\ld/a$ and (b) $\ldlam$ versus $P/\tai$ for $l/a = $~0.05, 0.1, 0.2, and 0.4, with $\mutildep = 0.8$. Symbols in panel (b) correspond to the analytical solution in the TT approximation given by Equation~(\ref{eq:ldlamboth}), while the vertical dotted line is the approximate transitional period given by Equation~(\ref{eq:pcrit}) for $l/a=0.1$. The shaded area denotes the range of observed periods of thread oscillations. \label{fig:res1}}
\end{figure}

 \begin{figure}[!htp]
\epsscale{0.49}
\plotone{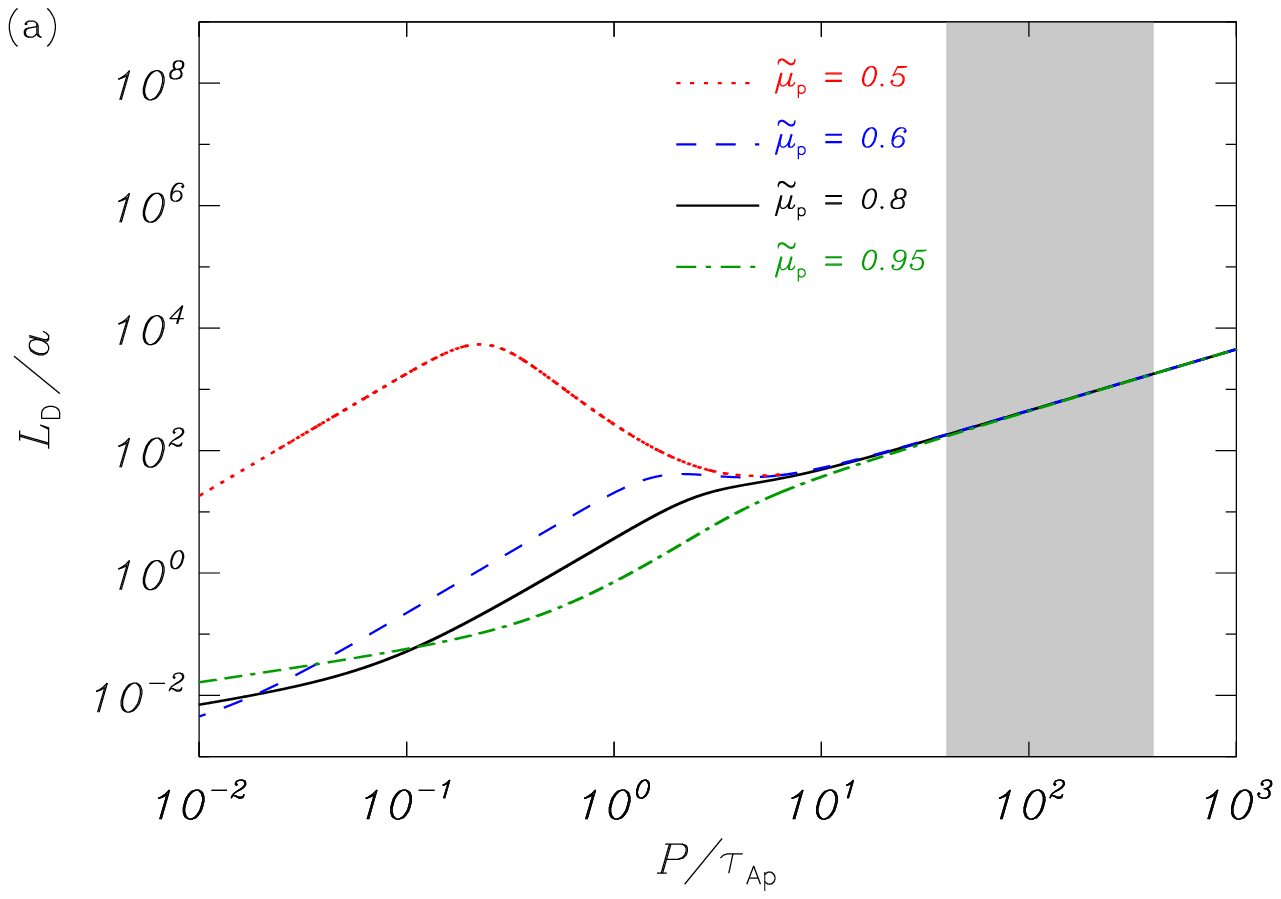}
\plotone{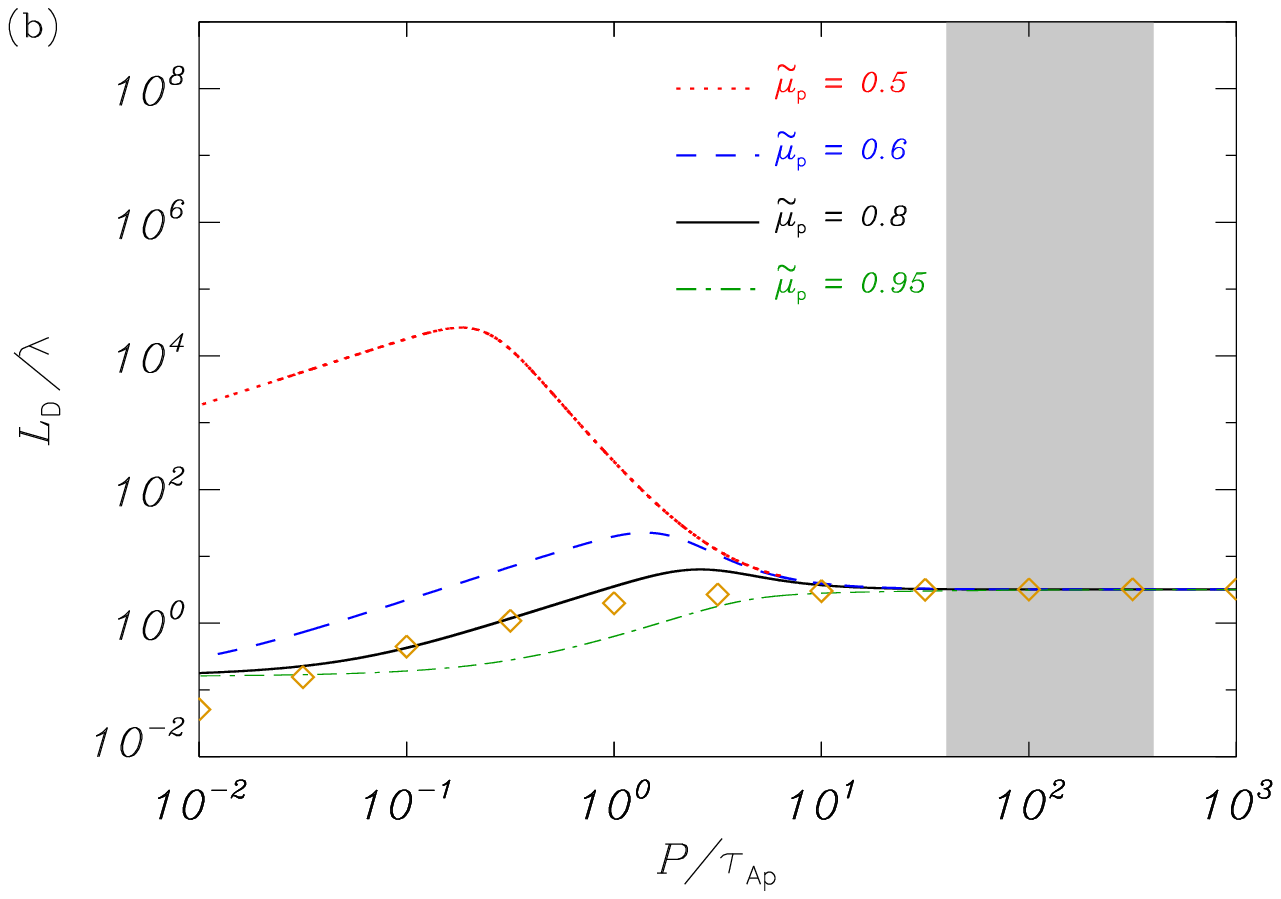}
\caption{Results for the kink mode spatial damping in the case $l/a \neq 0$: (a) $\ld/a$ and (b) $\ldlam$ versus $P/\tai$ for $\mutildep = $~0.5, 0.6, 0.8, and 0.95, with $l/a = 0.2$. Symbols in panel (b) correspond to the analytical solution in the TT approximation given by Equation~(\ref{eq:ldlamboth}). The shaded area denotes the range of observed periods of thread oscillations. \label{fig:res2}}
\end{figure}

Next, we study the general case $l/a \neq 0$. We adopt a sinusoidal density profile within the inhomogeneous transitional layer \citep{rudermanroberts}. As the Alfv\'en resonance position, $r_{\rm A}$, is needed for the computations of the resonant damping, we follow a two-step procedure. First, we solve the dispersion relation for a fixed $\omega$ in the case $l/a = 0$ and determine $k_{z \rm R}$. Then, we assume that the value of $k_{z \rm R}$ is approximately the same in the case $l/a \ne 0$, meaning that the resonant condition is $\omega = k_{z \rm R} v_{\rm A}\left( r_{\rm A} \right)$. In the case of a sinusoidal profile, the expression of the resonant position can be analytically obtained from the resonant condition as
\begin{equation}
 r_{\rm A} = a + \frac{l}{\pi} \arcsin \left( \frac{ \rhop + \rhoc}{ \rhop - \rhoc} - \frac{2 \vap^2 k_{z \rm R}^2}{\omega^2} \frac{\rhop}{\rhop-\rhoc} \right). \label{eq:ra}
\end{equation}
Finally, we compute $\left| \partial_r \rho  \right|_{r_{\rm A}}$ using the previously determined $r_{\rm A}$ by means of Equation~(\ref{eq:ra}) and solve the dispersion relation with these parameters to obtain the actual $k_{z \rm R}$ and $k_{z \rm I}$. Figure~\ref{fig:res1} shows the results of these computations for different values of $l/a$ when the ionization degree has been fixed to $\mutildep=0.8$, whereas Figure~\ref{fig:res2} displays the equivalent computations for different values of $\mutildep$ when the transverse inhomogeneity length scale has been fixed to $l/a=0.2$. Since the wavelength is not affected by the value of $l/a$ and has the same behavior as in Figure~\ref{fig:cow}(a), both Figures~\ref{fig:res1} and \ref{fig:res2} focus on $\ld/a$ and $\ldlam$. We obtain two different behaviors of the solutions depending on the period. For small $P/\tai$, the damping length is independent of $l/a$ and is governed by the value of $\mutildep$. On the contrary, for large $P/\tai$ the damping length depends on $l/a$ but is independent of $\mutildep$. This result indicates that resonant absorption dominates the damping for large $P/\tai$, whereas Cowling's diffusion is more relevant for small $P/\tai$. The approximate transitional period, namely $P_{\rm tr}$, in which the damping length by Cowling's diffusion becomes smaller than that due to resonant absorption can be estimated by setting $\epsilon \approx 1$ in Equation~(\ref{eq:delta}) and writing $P_{\rm tr} = 2\pi/\omega$. Then, one obtains
\begin{equation}
 P_{\rm tr} \approx 2\pi^2 \mathcal{F} \frac{a}{l} \frac{\etacb}{\ck^2 m}\frac{\rhop+\rhoc}{\rhop-\rhoc} = 2\pi^2 \mathcal{F} \frac{a}{l}  \frac{\etac}{\ck^2 m }\frac{\rhop}{\rhop-\rhoc}. \label{eq:pcrit}
\end{equation}
This transitional period is in good agreement with the numerical results (see the vertical dotted line in Figure~\ref{fig:res1}(b)). In addition, we see that $P_{\rm tr}$ is much smaller than the typically observed periods, indicating that resonant absorption is the dominant damping mechanism in the relevant range.

Finally, we check that the analytical approximation of $\ldlam$ given by Equation~(\ref{eq:ldlamboth}) provides an accurate description of the kink mode spatial damping in the relevant range of periods (compare the symbols and the solid lines in Figures~\ref{fig:res1}(b) and \ref{fig:res2}(b)).

\section{DISCUSSION AND CONCLUSION}
\label{sec:conclusions}

In this paper, we have studied the spatial damping of kink waves in prominence threads. Resonant absorption and Cowling's diffusion are the damping mechanisms taken into account. Both analytical expressions and numerical results indicate that, in the range of typically observed periods of prominence thread oscillations, the effect of Cowling's diffusion (and so the ionization degree) is negligible. On the other hand, resonant absorption provides an efficient damping in agreement with the study of \citet{spatialterradas} in the context of coronal loop oscillations. These conclusions are equivalent to those obtained by \citet{solerRAPI} in the case of temporal damping.

We point out that small values of $\ldlam$ are obtained by resonant absorption in the observationally relevant range of periods, which is consistent with the reported strong damping of the oscillations. The analytical estimation of $\ldlam$ given by Equation~(\ref{eq:ldlamboth}) is very accurate in the observationally relevant range of periods, and the contribution of Cowling's diffusion can be dropped from Equation~(\ref{eq:ldlamboth}) because the plasma ionization degree turns out to be irrelevant for the damping. Therefore, for kink modes ($m=1$) the radio $\ldlam$ simplifies to
\begin{equation}
 \frac{\ld}{\lambda} \approx \mathcal{F} \frac{a}{l} \frac{\rhop+\rhoc}{\rhop-\rhoc}, \label{eq:taulamfin}
\end{equation}
which coincides with the expression provided by \citet{spatialterradas}. As $\frac{\rhop+\rhoc}{\rhop-\rhoc} \to 1$ for typical prominence and coronal densities, this factor can be dropped from Equation~(\ref{eq:taulamfin}), meaning that the ratio $\ldlam$ depends almost exclusively on the transverse inhomogeneity length scale, $l/a$, and the form of the density profile through $\mathcal{F}$ as
\begin{equation}
 \frac{\ld}{\lambda} \approx \mathcal{F} \frac{a}{l}. \label{eq:taulamfin2}
\end{equation}
In the case of coronal loop oscillations studied by \citet{spatialterradas}, the factor $\frac{\rhop+\rhoc}{\rhop-\rhoc}$ cannot be dropped from their expressions, meaning that in coronal loops the ratio $\ldlam$ significantly depends on the density contrast. Therefore, information about the parameters $l/a$ and $\mathcal{F}$ in prominence threads could be determined by using Equation~(\ref{eq:taulamfin2}) along with accurate measurements of the damping length and the wavelength  provided from the observations. However, since the precise form of the transverse density profile in prominence threads is unknown, we have to assume an ad hoc profile, i.e., a value of $\mathcal{F}$, to infer the transverse inhomogeneity length scale from the observations, which can introduce some uncertainties in the estimation of $l/a$. 

For example, let us assume that the ratio $\ldlam$ has been determined from an observation of damped kink waves in a prominence thread and we want to compute the transverse inhomogeneity length scale of the thread. For simplicity, we consider that the transverse density profile in the inhomogeneous layer is either linear or sinusoidal. Denoting as  $\left( l/a \right)_{\rm lin}$ the value of $l/a$ computed assuming a linear profile, and $\left( l/a \right)_{\rm sin}$ the corresponding value for a sinusoidal profile, the relation between both of them is
\begin{equation}
 \frac{\left( l/a \right)_{\rm lin}}{\left( l/a \right)_{\rm sin}} = \frac{\pi}{2} \approx 1.57,
\end{equation}
pointing out that the relative uncertainty of $l/a$ is larger than 50\%, and the inaccuracy could be even larger if other profiles are considered. This fact should be taken into account in future seismological determinations of this parameter.

The present investigation is a first step for the study of the spatial damping of kink waves in prominence fine structures. Here, we have adopted a simple model of a prominence thread. Some effects that might influence the kink mode propagation and damping are not included in the present paper. Among them, plasma inhomogeneity along the thread may affect somehow the amplitude of a propagating kink mode, whereas the presence of flows affects the damping by resonant absorption \citep[see][]{terradasflow}. The influence of these and other effects will be the subject of forthcoming works.

\begin{acknowledgements}
      The authors acknowledge the financial support received from the Spanish MICINN and FEDER funds (AYA2006-07637). The authors also acknowledge discussion within ISSI Team on Solar Prominence Formation and Equilibrium: New data, new models. RS thanks the CAIB for a fellowship.
\end{acknowledgements}

\end{document}